\documentclass[fleqn,usenatbib]{rasti}

\usepackage{newtxtext,newtxmath}
\usepackage[T1]{fontenc}

\DeclareRobustCommand{\VAN}[3]{#2}
\let\VANthebibliography\thebibliography
\def\thebibliography{\DeclareRobustCommand{\VAN}[3]{##3}\VANthebibliography}

\usepackage{graphicx}	% Including figure files
\usepackage{amsmath}	% Advanced maths commands
\usepackage{orcidlink}
\usepackage{}

\usepackage{textcomp}

\title[MALLORN]{MALLORN: Many Artificial LSST Lightcurves based on Observations of Real Nuclear transients}

\author[D. Magill et al.]{
Dylan Magill$^{1}$\orcidlink{0009-0000-6521-8842},
M. Nicholl$^{1}$\orcidlink{0000-0002-2555-3192},
V. Anilkumar$^{2}$\orcidlink{0009-0008-3146-287X},
S. van Velzen$^{2}$\orcidlink{0000-0002-3859-8074},
X. Sheng$^{1}$\orcidlink{0000-0002-6527-1368},
\newauthor
T. S. Mai$^{3}$\orcidlink{0000-0003-4599-1525},
H. V. Tran$^{3}$\orcidlink{0000-0003-2690-1379},
N. P. Doan$^{3}$\orcidlink{0009-0004-2552-3681},
T. Moore$^{1,4}$\orcidlink{0000-0001-8385-3727},
S. Srivastav$^{5}$\orcidlink{0000-0003-4524-6883},
\newauthor
D. R. Young$^{1}$\orcidlink{0000-0002-1229-2499},
C. R. Angus$^{1}$\orcidlink{0000-0002-4269-7999},
J. Weston$^{1}$\orcidlink{0009-0002-9460-9900}
\\
$^{1}$Astrophysics Research Centre, School of Mathematics and Physics, Queen's University Belfast, Belfast BT7 1NN, UK\\
$^{2}$Leiden Observatory, Leiden University, Postbus 9513, NL-2300 RA Leiden, the Netherlands \\
$^{3}$School of Electronics, Electrical Engineering and Computer Science, Queen’s University Belfast, Belfast BT9 5BN, UK  \\
$^{4}$Space Telescope Science Institute, 3700 San Martin Dr., Baltimore, MD21218, USA \\
$^{5}$Astrophysics sub-Department, Department of Physics, University of Oxford, Keble Road, Oxford, OX1 3RH, UK \\
}

\date{Accepted XXX. Received YYY; in original form ZZZ}

\pubyear{2025}

\begin{document}
\label{firstpage}
\pagerange{\pageref{firstpage}--\pageref{lastpage}}
\maketitle

\begin{abstract}
The Vera C. Rubin Observatory's 10-Year Legacy Survey of Space and Time (LSST), is expected to produce a hundredfold increase in the number of transients we observe. However, there are insufficient spectroscopic resources to follow up on all of the wealth of targets that LSST will provide. As such it is necessary to be able to prioritise objects for followup observations or inclusion in sample studies based purely on their LSST photometry. We are particularly keen to identify tidal disruption events (TDEs) with LSST. TDEs are immensely useful for determining black hole parameters and probing our understanding of accretion physics. To assist in these efforts, we present the Many Artificial LSST Lightcurves based on the Observations of Real Nuclear transients (MALLORN) data set and the corresponding classifier challenge for identifying TDEs. 
MALLORN comprises 10178 simulated LSST light curves, constructed from real Zwicky Transient Facility (ZTF) observations of 64 TDEs, 727 nuclear supernovae and 1407 AGN with spectroscopic labels using Gaussian process fitting, empirically-motivated spectral energy distributions from \texttt{SNCosmo} and the baseline from the Rubin Survey Simulator. Our novel approach can be easily adapted to simulate transients for any photometric survey using observations from another, requiring only the limiting magnitudes and an estimate of the cadence of observations. The MALLORN Astronomical Classification Challenge, launched on Kaggle on 15/10/2025, will allow competitors to test their photometric classifiers on simulated LSST data to find TDEs and improve upon their capabilities prior to the start of LSST.

%This is a simple template for authors to write new RASTI papers.
%The abstract should briefly describe the aims, methods, and main results of the paper.
%It should be a single paragraph not more than 250 words. - currently 248 words
%No references should appear in the abstract.
\end{abstract}

% Include between one and six keywords.

%%%%%%%%%%%%%%%%%%%%%%%%%%%%%%%%%%%%%%%%%%%%%%%%%%

%%%%%%%%%%%%%%%%% BODY OF PAPER %%%%%%%%%%%%%%%%%%

\section{Introduction}
\label{introduction}

The Vera C. Rubin Observatory's 10-Year Legacy Survey of Space and Time \citep[LSST;][]{LSST} is expected to revolutionise time-domain astronomy. It is predicted to produce a hundredfold increase in the number of transients (such as supernovae and tidal disruption events) that we are able to discover, thanks to its large mirror size (8.4m) allowing it to detect fainter and more distant transients compared to previous wide-field survey telescopes.

However, we have insufficient spectroscopic resources to follow up on every photometrically discovered transient that LSST is able to identify, even with dedicated spectroscopic follow-up from facilities including ESO 4MOST, operating the Time Domain Extragalactic Survey \citep[TiDES;][]{TiDES}, the Son of X-shooter \citep[SOXS;][]{SOXS} instrument on the New Technology Telescope (NTT), and US NOIRLab telescopes such as Gemini \citep[][]{Gemini} and SOAR \citep[][]{SOAR}. As spectroscopic observations require much longer exposure times compared to photometry, and most spectrographs can only focus on one object at a time, we will struggle to get spectra for even a few percent of the objects that LSST will discover.  Therefore, in order to benefit from the wealth of targets that LSST will provide, it will be necessary to identify which objects to prioritise for follow-up observations and to conduct analysis of larger samples based on photometry alone.

Selection of targets based on photometry is particularly important for identifying examples of rare transient classes. One rare transient of great physical interest are tidal disruption events (TDEs). TDEs occur whenever an unfortunate star is ripped apart by the immense gravitational forces it experiences if it is scattered onto an orbit that passes too close to a supermassive black hole \citep[][]{Gezari2021}. TDEs are a relatively recent astronomical discovery, first predicted in the 1970s \citep[][]{Hills1975, Lidskii1979} and first observed in optical wavelengths in the early 2010s \citep[][]{VanVelzen2011, Gezari2012}.

TDEs are of key astrophysical significance because they can be used to identify the mass of the disrupting supermassive black hole (SMBH). This can be be applied to smaller SMBHs which may not otherwise be detectable for a dynamical approach. This allows us to investigate scaling relations and black hole growth at the lower mass end of the SMBH mass function \citep[][]{Ramsden2025}. They are also sensitive probes of galaxy dynamics/evolution \citep[][]{French2020}, and accretion physics \citep[e.g.][]{Guolo2025}, and may be detectable to very high redshifts due to their high luminosity in the ultraviolet \citep[][]{Karmen2025}.

Our understanding of TDEs is currently limited by the small sample size of observed objects (\textasciitilde100). Fortunately, it has been predicted that LSST will detect roughly a few thousand TDEs per year \citep[][]{VanVelzen2011,Bricman2020, BucarBricman2023}. The significant uncertainty in this number is due to the uncertainty in SMBH mass distribution, which the volumetric TDE rate is heavily dependent on. Additionally, the fraction of these TDEs that is actually identified is limited by the availability of spectroscopic follow-up resources to classify discovered transients.

Machine learning classifiers are now being utilised to photometrically select TDEs. The most prominent current TDE classifiers are the NEural Engine for Discovering Luminous Events \citep[NEEDLE;][]{Sheng2024}, TDEscore \citep[][]{Stein2024}, Finding Luminous Exotic Extragalactic Transients \citep[FLEET;][]{Gomez2020,Gomez2023} and the Automatic Learning for the Rapid CLassification of Events (ALeRCE) lightcurve classifier \citep[][]{Pavez-Herrera2025}. These classifiers typically select for blue transients in the nuclei of centrally concentrated galaxies. The most common contaminants in the search for TDEs are active galactic nuclei (AGN) and Type Ia supernovae, which make up the bulk of the observed nuclear transients \citep[][]{Dgany2023, Hlozek2023}.

To evaluate the performance of multiple classifiers and to allow for them to improve upon their capabilities, the scientific community has developed the concept of classification challenges for classifier algorithms. The data set provided to the competitor is split into two parts: a training set and a testing set. The competitor is able to train their classifier on the training portion of the data set in which the types of objects are known, and will then use their classifier to attempt to identify the objects in the testing set. The competitor will then submit their results to the organisers of the classification challenge and have their performance evaluated. 

Recent astronomical classification challenges include the Photometric LSST Astronomical Time-Series Classification Challenge \citep[PLAsTiCC;][]{Kessler2019, Malz2019, Hlozek2023} and the Extended LSST Astronomical Time-Series Classification Challenge \citep[ELAsTiCC;][]{ELAsTiCC}. PLAsTiCC created 3.5 million simulated objects for LSST using SNANA \citep[][]{Kessler2009} and asked competitors to provide likelihoods for each object of belonging to each transient class. PLAsTiCC focused on producing successful classifications for all objects in order to equally weight the various science goals of LSST. ELAsTiCC was the successor competition that converted the data into a stream of alerts fed into the LSST brokers (platforms to host and add value to real-time detections of transient and variable objects) to allow for classifier testing on those platforms. Additional examples of previous classifier challenges for other science goals include the Supernova Photometric Classification Challenge \citep[SNPCC;][]{SNPCC}, Mapping Dark Matter \citep[][]{Kitching2011}, Observing Dark Worlds \citep[][]{Harvey2014} and Galaxy Challenge \citep[][]{Dieleman2015}.

PLAsTiCC and ELAsTiCC both used models to simulate the data, some of which were based on coarse analytic approximations, and so may not be fully accurate representations of real transients. Training ML algorithms on models rather than data therefore runs the risk of resulting in poor performance when deployed in the real world.  

To aid in the preparation for LSST, we have produced the Many Artificial LSST Lightcurves based on the Observations of Real Nuclear transients (MALLORN) data set. This is a simulated photometric data set of nuclear transients for LSST, with realistic cadence and luminosities, that has been developed using real observations from the Zwicky Transient Facility \citep[ZTF;][]{ZTF}. This data set is being shared as a classifier challenge to allow others in the field to test and improve upon their photometric classifiers. This will allow the scientific community to better prepare for the start of LSST and improve upon our collective ability to effectively use the wealth of data that LSST will provide.

In this paper we describe the method used to create simulated LSST lightcurves, the construction of the MALLORN data set and what we have released in the classification challenge\footnote{\url{https://www.kaggle.com/competitions/mallorn-astronomical-classification-challenge/overview}}. 
%We also provide Google Colab notebooks which show the general method used to create a simulated LSST lightcurve\footnote{MALLORN Data Production Notebook - \url{https://colab.research.google.com/drive/1oy96r29Zs4U5Hl-THsZPCnOQbuz21hl5}} and explain how to interact with the MALLORN data\footnote{Using MALLORN Data Notebook - \url{https://colab.research.google.com/drive/1N7Q1bxc2gxBuOv2eD3fTYsrxoLC2dQAP}}. 

%This is a simple template for authors to write new RASTI papers.
%See \texttt{rasti\_guide.tex}
%for a full user guide.

%All papers should start with an Introduction section, which sets the work
%in context, cites relevant earlier studies in the field by \citet{Fournier1901},
%and describes the problem the authors aim to solve \citep[e.g.][]{vanDijk1902}.
%Multiple citations can be joined in a simple way like \citet{deLaguarde1903, delaGuarde1904}.

\section{Simulated Lightcurve Production}

This section will detail the process of producing a simulated LSST lightcurve from a real ZTF lightcurve. More details on the production at scale and specific choices made for each class of object are provided in Section \ref{LSST_Dataset_production}. For MALLORN, this method used specific data from ZTF and specific survey characteristics appropriate to LSST. However, provided you have the limiting magnitudes of the instrument and an estimate of the cadence for the target survey, this method should be applicable to produce simulated data for any new photometric survey. 

The MALLORN Data Production Google Colab Notebook\footnote{\url{https://colab.research.google.com/drive/1oy96r29Zs4U5Hl-THsZPCnOQbuz21hl5}} details the process of going from a real ZTF lightcurve to a simulated LSST lightcurve. The methods are outlined as clearly as possible to ease the process for anyone who wishes to replicate this method for any future use.

\subsection{Initial lightcurve selection}
\label{init_lc_selection}

The data set used to produce MALLORN consisted of all of the ZTF forced photometry of nuclear transients from 2017-2023. This sample was created by applying a filter to \texttt{AMPEL} archive of ZTF alerts \citep{AMPEL}. This so-called Nuclear Filter was originally written and employed by \citet{VanVelzen2019,VanVelzen2021,Hammerstein2021} to select nuclear transients in real-time from the ZTF alert stream.  

To pass the Nuclear Filter an object must have a high likelihood of a galactic host, at least one detection brighter than 20 mag and an angular distance to the core of less than 0.5 arcseconds. For more information on this filter, please see \citet[][]{Reusch2024}. The sample produced by this filter contained a total of 11190 objects.

The data set was crossmatched with the \texttt{Transient Name Server}\footnote{\url{https://www.wis-tns.org/}} using the ZTF object names and coordinates (RA and Dec) to determine the spectroscopic type and redshift where available. This produced 791 objects for which the type and redshift were known. Further manual crossmatching was able to identify an additional 38 TDEs within the ZTF data set, bringing the total number of confirmed TDEs included to 64. This sample contains both TDEs in galaxies without AGN and TDEs in galaxies with AGN. Further crossmatching with version 8 of the \texttt{Milliquas} catalogue \citep[][]{milliquas} identified 1407 additional AGN within the ZTF data set. There were some objects misclassified as Type II superluminous supernovae on \texttt{TNS} that were corrected to AGN based on the contaminants identified in \citet{Pessi2025}.

We first run the ZTF forced photometry pipeline for all transients in our sample. This returns the time, count rate and zeropoint for each observation, in the ZTF $g$ and $r$ bands. We use the zeropoint to convert each observation to a flux density measurement in microjanskys ($\mu$Jy).

Following the approach outlined in \citep{Hammerstein2021}, a baseline correction is carried out on the lightcurve. We define the baseline as the inverse-variance weighted average of the flux a hundred days before the rise of the transient event. This correction step subtracts this average value from the complete lightcurve and propagates the errors, bringing the mean of the pre-transient event flux to zero. In addition, the reduced $\chi^2$ values of the baseline are brought to one by scaling the uncertainties of the flux measurement. 

We next correct for Galactic extinction using the \texttt{dustmaps} \citep[][]{dustmaps} and \texttt{extinction} \citep[][]{extinction} packages. The extinction coefficient, $E(B-V)$, for the specified RA and Dec is determined using the Schlafly \& Finkbeiner dustmap \citep[][]{S_F_dustmap}. Using the $E(B-V)$ value and the effective wavelength for the respective filter ($g$ or $r$ band) sourced from the Spanish Virtual Observatory filter service \citep[][]{SVO}, the extinction at a given wavelength ($A_\lambda$) can then be calculated using the extinction laws from \citet{Fitzpatrick1999} and assuming a $R_V = 3.1$ for Milky Way extinction \citep[][]{Cardelli1989}. The de-extincted flux values ($F_\lambda$) can then be calculated from the initial extincted flux ($F_{\lambda,\text{Ext}}$) following:

\begin{center}
    
    \begin{equation}
        F_\lambda = F_{\lambda,\text{Ext}}  10^{\frac{A_\lambda}{2.5}}
    \end{equation}
\end{center}

\subsection{Gaussian Process Fit}
\label{GP_fit}

We then take the baseline- and extinction-corrected lightcurves, and select the band ($g$ or $r$) with the most observations. The peak time for the selected band is found by determining the time which corresponds to the greatest flux value. A Gaussian Process (GP) fit is then applied to the data for the selected band. This produces a continuous lightcurve, interpolating between any gaps in the observations. There is strong precedent in the literature for the application of Gaussian process fitting to the photometry for transient objects \citep[][]{ForemanMackey2017,Boone2019,Kornilov2023}, though see \citet[][]{Stevance2023} for limitations of lightcurve GP fitting.

\begin{figure}
	\includegraphics[width=\columnwidth]{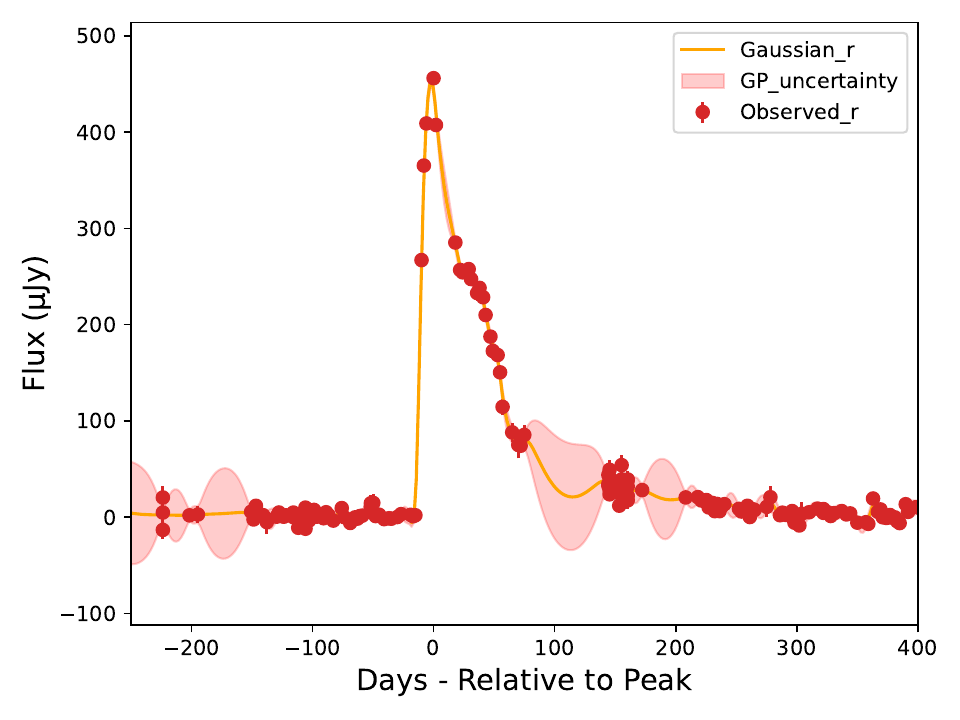}
    \caption{Plot of flux ($\mu$Jy) against time (relative to peak) showing the r-band observations for SN II ZTF22abcfics/SN2022ryv (red) and the GP fit to those observations (orange). The GP fit reliably follows the trend within the data points. During large gaps in observations the uncertainty in the GP fit is inflated. We show the GP uncertainty only for visualisation and use a different method to determine the uncertainty in our simulated LSST observations.}
    \label{fig:gp_plot}
\end{figure}

The kernel used for the GP fit consists of three components combined as: 
\begin{equation}
    K = K_C \times K_M + K_{WN}
\label{kernels}
\end{equation}
The first is a constant kernel ($K_C$), which defines amplitude and range of possible parameters. Another component is the Mat\'ern kernel ($K_M$), which is a generalised radial basis function (RBF) kernel - parameterised by a given length scale and used to determine the similarity of two data points. This sets the characteristic timescale over which the flux changes. The advantage of a Mat\'ern kernel over a standard RBF kernel is that it contains an additional smoothness parameter which allows for more efficient computation. The final component is a white noise kernel ($K_{WN}$), which accounts for possible variation due to noise in the data. An example of a GP fit to a ZTF lightcurve is shown in Figure \ref{fig:gp_plot}.

To verify the validity of the GP fit, we calculate the mean squared error (MSE) of the fit. To do this, the flux is first normalised so that the amplitude ranges from 0 to 1 and then a set fraction of data points are randomly removed. The GP fit is then re-ran and used to create simulated points at the positions of the removed points. The MSE is computed from the differences between the original points and the simulated points \citep[][]{Benidis2020}. An MSE below 0.2 is generally considered a good fit \citep[][]{Hastie2009}. Running on a test of 200 randomly selected objects, the GP fits returned a median value of 0.14. We don't allow any fits with an MSE greater than 0.3 to pass. Therefore, our value indicates that the GP fit produces an accurate reproduction of the full lightcurve.

Using a multidimensional GP fit which considered data from both bands to fit to one band was also tested. This produced little to no improvement to the fit, but did produce a measurable increase in computation time, which becomes significant when producing the full data set. Therefore, as this alternative produced little to no improvement at the expense of greater computational time, we chose to apply the GP fit only to the band with the most observations, with the other band later being used to verify the fits (see Section \ref{SNCosmo}).

\begin{figure}
	\includegraphics[width=\columnwidth]{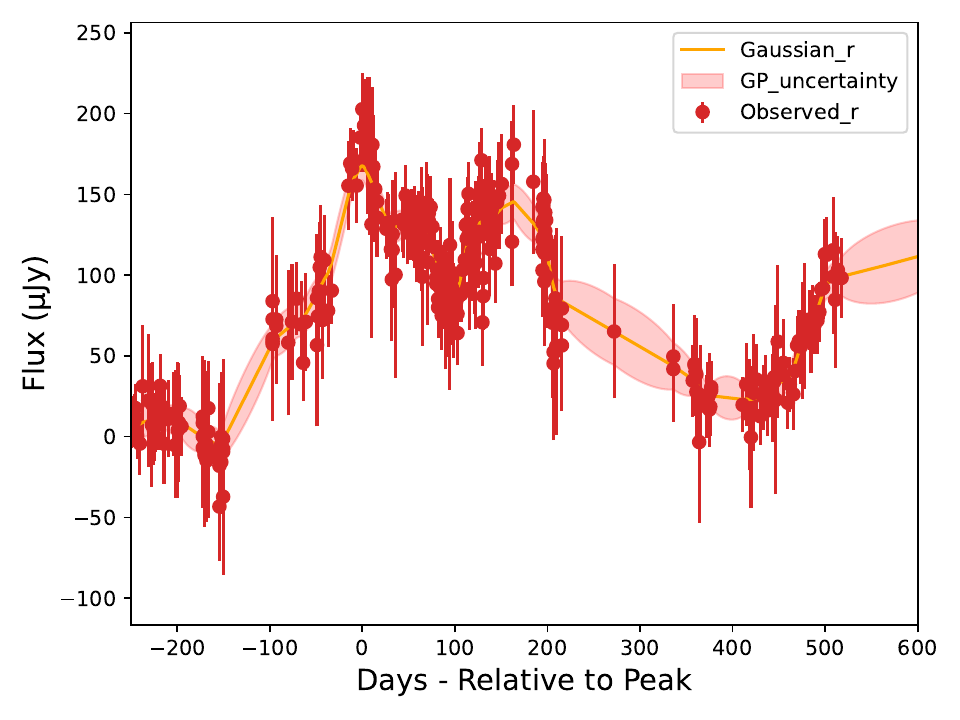}
    \caption{Plot showing the Damped Random Walk (DRW) fit to the observed AGN ZTF lightcurve (ZTF20acxfcmr). The DRW more effectively handles the stochasticity of an AGN lightcurve. During large gaps in observations the uncertainty in the GP fit is inflated. We show the GP uncertainty only for visualisation and use a different method to determine the uncertainty in our simulated LSST observations.}
    \label{fig:DRW_fit}
\end{figure}

In order to more appropriately handle the stochastic nature of AGN lightcurves, a damped random walk (DRW) time-series model was used to produce a continuous fit to these data. A DRW can be defined by an exponential covariance matrix (kernel) and two main parameters: the long-term variability amplitude, which defines the largest possible fluctuation range of a lightcurve, and the signal decorrelation timescale, which defines the minimum timescale beyond which the luminosity differences over increasing time intervals are no longer correlated with the length of those intervals. There are multiple examples in the literature of DRWs being used to model the variability in the lightcurves of quasars and AGN  \citep[][]{Kelly2009,MacLeod2010,Ivezic_agn}. The approach used here is based on that of \citet{Sheng2022}, which employs the \texttt{EzTao} toolkit \citep[][]{EzTao}. \texttt{EzTao} is an efficient package for simulating, modelling and recovering the parameters of DRW and higher-order time-series models for AGN variability, implemented using Gaussian processes. Solely the DRW kernel is then used in a GP fit to produce a continuous lightcurve for the AGN. An example of this fit is shown in Figure \ref{fig:DRW_fit}.

Our Gaussian process fit yields inflated uncertainties, particularly during the gaps between observations. This behaviour arises from the use of a zero-mean function with a stationary kernel, which models the transient as a deviation from the mean (Anilkumar et al., in prep.). As shown in Figure \ref{fig:gp_plot}, these effects are most pronounced further away from the transient peak. However, this does not impact our subsequent analysis, as data point errors are determined using the \texttt{Rubin Survey Simulator} (see Section \ref{2.5}) rather than the GP uncertainty.

\subsection{Rescaling luminosity for simulated lightcurve}
\label{rescaling}

Since LSST observes to a deeper limiting magnitude than ZTF (e.g.~in $r$-band, $m_{{\rm lim,ZTF}}$ = 20.6, $m_{{\rm lim,LSST}}$ = 24.7), a transient at a given redshift would be detected with high confidence in LSST both earlier and later in time than the same transient in ZTF. Simulating this would require an extrapolation in time, which would require a more complicated model not anchored in the ZTF data. 

To avoid this issue, we instead place each simulated lightcurve at a distance where the signal-to-noise in LSST would approximately match the observed signal-to-noise ratio of the observed lightcurve in ZTF. Effectively, this shifts the simulated population to higher redshifts where the bulk of LSST transients will be detected, while also preserving the shape of the observed luminosity distribution from ZTF. We note that this will result in few low-redshift transients in our simulated dataset; however such objects are likely to be identifiable already with existing methods and surveys, complemented by the additional colour information from LSST. It is for the more distant objects that we need to improve lightcurve classifiers.

The real distance to each ZTF object is calculated from its known redshift, using \texttt{astropy.cosmology} with \texttt{WMAP9} from \citet[][]{WMAP9}. The following equation is then used to calculate a simulated distance at which the observed flux from the object is at the same factor above the LSST detection limit as the original flux was above the ZTF detection limit:

\begin{equation}
    D_{\text{LSST}} = D_{\text{ZTF}} \times \sqrt{\frac{F_{\text{lim, ZTF}}}{F_{\text{lim, LSST}}}} \times \sigma
\end{equation}

A random scatter component ($\sigma$ = 0.9-1.1) is included within this equation to enhance the variety of simulated objects produced for the data set. This simulated LSST distance for each object is then used to determine the simulated LSST redshift using the same cosmology calculator. The simulated flux is rescaled following:

\begin{equation}
    F_{\text{LSST}} = F_\text{ZTF} \times \frac{F_{\text{lim, ZTF}}}{F_{\text{lim, LSST}}} \times \delta
\end{equation}

To further enhance the variety of simulated objects produced for the data set, a scatter component ($\delta$) is selected from a normal distribution, which can be tuned differently if desired to represent the diversity inherent to different transient classes. For discussion on the values used in our data challenge, see Section \ref{Large Scale Production}.

Finally, the effect of time dilation on the observed lightcurve was taken into account using the known ZTF redshift ($z_{\text{ZTF}}$) and the simulated LSST redshift ($z_{\text{LSST}}$):

\begin{equation}
\label{eq:redshift}
    (t-t_\text{peak})_\text{LSST} = (t-t_\text{peak})_\text{ZTF} \times \frac{1 + z_{\text{LSST}}}{1 + z_{\text{ZTF}}}
\end{equation}

\subsection{Using SNCosmo models to create data for missing bands}
\label{SNCosmo}
One of the main challenges in simulating LSST data from ZTF observations is that ZTF observes predominantly in two bands ($g$ \& $r$ -- though with occasional $i$-band observations), whereas LSST will observe in six bands ($u$, $g$, $r$, $i$, $z$ \& $y$). Therefore, it is necessary to simulate data for the other four bands not observed by ZTF, and important that this is done in such a way as to still be grounded in the real transient observations. Since the surveys share some bands, we will henceforth differentiate observations/simulations in ZTF and LSST using the subscripts `Z' and `L'.

To accomplish this, we used the \texttt{SNCosmo} package \citep[][]{sncosmo}, which allows us to generate empirically-motivated spectral energy distribution (SED) models for different types of transients at a given redshift and time since explosion. For each ZTF transient with an available spectroscopic type, we select an appropriate \texttt{SNCosmo} model, and use it to generate two time-dependent SEDs -- one at the observed redshift and one at the simulated LSST redshift. 

From these time-dependent SED models, we can then compute the colour evolution and k-correction in each LSST band with respect to our reference band ($g_{\rm Z}$ or $r_{\rm Z}$); i.e. $r_{\rm Z}-u_{\rm L}$, $r_{\rm Z}-g_{\rm L}$, $r_{\rm Z}-r_{\rm L}$, etc. These colour curves can then be added to the (time-dilated) GP fit of the reference band to produce k-corrected simulated lightcurves for all six LSST bands using:

\begin{equation}
    F_{\rm L}(t'') = F_{\rm Z}(t')  10^{(m_{\rm Z}(t') - m_{\rm L}(t'') )/ 2.5}.
    \label{eq:colour_flux}
\end{equation}

$F_{\rm Z}$ and $F_{\rm L}$ are the fluxes in the ZTF reference band and the simulated LSST band respectively. $m_{\rm Z} - m_{\rm L}$ is the colour computed with SNCosmo. All of the variables are functions of time, with $t'=(t-t_{\rm peak})_{\rm ZTF}$ and $t''=(t-t_{\rm peak})_{\rm LSST}$; these quantities are related through equation \ref{eq:redshift}.

These colour curves are applied only in a restricted temporal region around the transient. This is necessary as it prevents the colour curves from exaggerating random noise in ZTF around zero flux, which may produce a `detection' in another band if the colour term is large enough. The colour curves are set to zero outside of this range so that each curve follows that of the reference band. This is a reasonable approximation as we would expect photometry in all bands to fall within a standard deviation of the baseline before the transient has occurred and after it has faded. For most transient types, we define this range as $50 \text{ days} \times (1 + z)$ prior to the peak to $150 \text{ days} \times (1 + z)$ after the peak. For particularly long-lived transients this range is extended. For SN Ic-BL, SN IIn and SLSN, the defined range is  $100 \text{ days} \times (1 + z)$ prior to the peak to $300 \text{ days} \times (1 + z)$ after the peak. For TDEs, the defined range is $100 \text{ days} \times (1 + z)$ prior to the peak to $400 \text{ days} \times (1 + z)$ after the peak. The range limit is not applied to AGN lightcurves, as they are expected to vary continuously and as such may have colour evolution throughout their entire lightcurve.

\begin{figure}
	\includegraphics[width=\columnwidth]{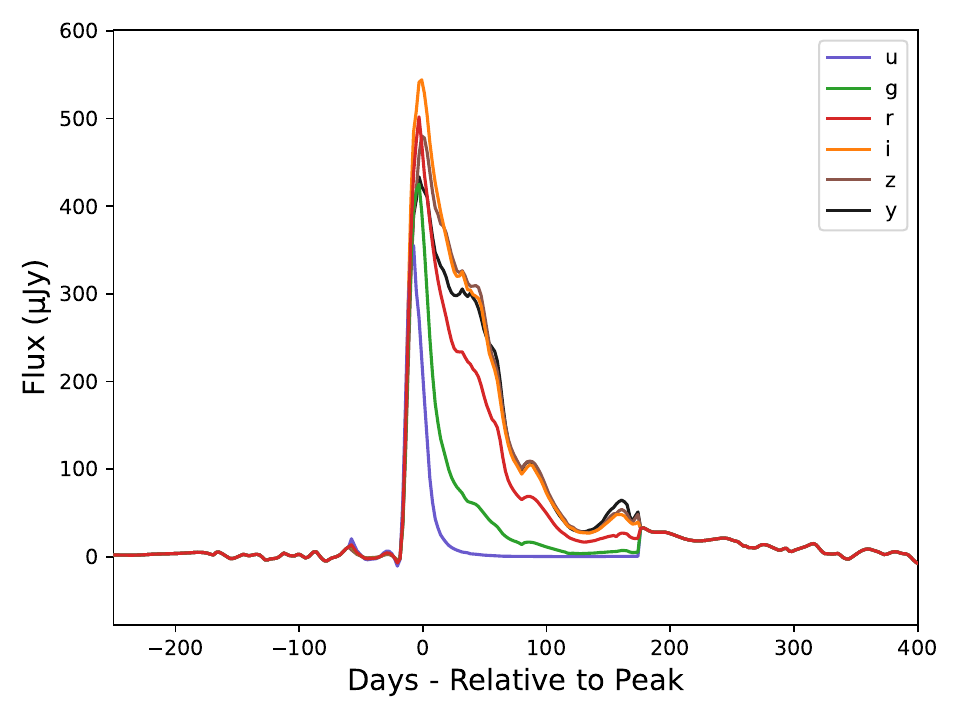}
    \caption{Six band plot of continuous k-corrected lightcurves for the SN II ZTF22abcfics/SN2022ryv. The $r$-band is produced via a GP fit of the observed data, whilst the other bands are produced using the colour differences generated from the \texttt{SNCosmo} models. The instances of more erratic variation in the $z$ and $y$ bands are the result of less data for the model for those bands, but are well within the bounds of the error values for their respective bands.}
    \label{fig:snc}
\end{figure}

This process therefore allows us to produce a full six-band continuous $k$-corrected lightcurve based on a real ZTF lightcurve and realistic SEDs for a given transient type. An example is shown in Figure \ref{fig:snc}.

\begin{figure}
	\includegraphics[width=\columnwidth]{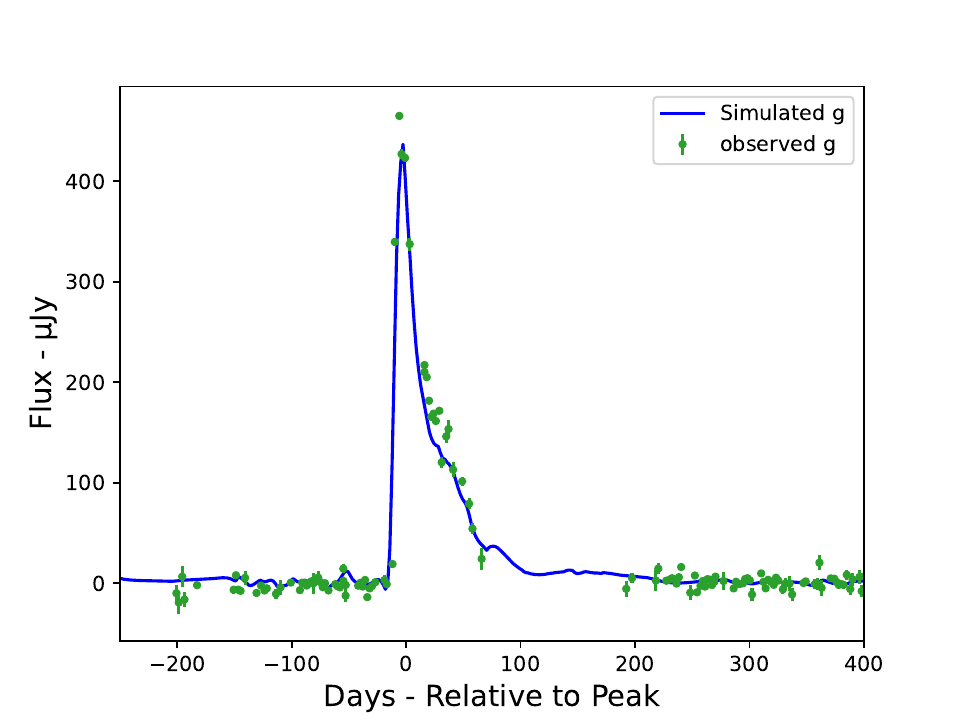}
    \caption{Plot comparing the simulated continuous $g$ band lightcurve generated via the observed redshift \texttt{SNCosmo} colour differences to the observed g-band data for the SN II ZTF22abcfics/SN2022ryv. The simulated data is shown to provide a reasonable match to the observed data, verifying the validity of this approach.}
    \label{fig:snc_vs_obs}
\end{figure}

To verify the validity of this approach, we also simulate data in the ZTF band with fewer data points (i.e.~the band not used for the GP fit) at the ZTF (true) redshift, and compare to the real observations for that band. As shown in Figure \ref{fig:snc_vs_obs}, the simulated continuous g-band lightcurve provides an acceptable match to the real ZTF g-band observations.

\subsection{Implementing LSST cadence}
\label{2.5}
In order to create realistic and representative simulated transients for LSST, we sample our model lightcurves using the LSST cadence \texttt{Baseline v5.0.0} from the \texttt{Rubin Survey Simulator} \citep[][]{rubin_sim, rubin_scheduler, rubin_baseline}.

For each model lightcurve, a random RA and Dec are chosen within the LSST footprint, and a random date for the lightcurve peak is chosen within the 10 year survey period. The cadence simulation is then used to determine the visit dates for each filter. These visit dates can then be matched with the corresponding values from the continuous lightcurves for each band.  

Often, these random dates fall in a period of solar conjunction, or when the field is in the `off-season' of the LSST rolling cadence. To avoid generating useless lightcurves which would be egregiously difficult to classify, we apply safety checks to ensure that the transient is actually detected in the simulation. We define a detection as having 10 data points across all bands with a signal-to-noise ratio $>5\sigma$. If the transient is not detected, the matching is repeated with newly generated co-ordinates and date.

We then apply Galactic extinction to the simulated LSST lightcurves. This is carried out in an inverse of the method described in Section \ref{init_lc_selection}. The effective wavelengths used to compute the extinction for each LSST band were: $u$ = 3641\AA, $g$ = 4704\AA, $r$ = 6155\AA, $i$ = 7504\AA, $z$ = 8695\AA \space \& $y$ = 10056\AA. These values were sourced from the SVO Filter Service \citep[][]{SVO}.

The \texttt{Rubin Survey Simulator} baseline contains $5\sigma$ detection limits for each observation. Given the faintness of our targets, we assume that all measurements will be sky-noise limited. We therefore estimate the errors on each photometric point by converting these limits from AB magnitude to flux and multiplying by 0.2.

\begin{figure}
	\includegraphics[width=\columnwidth]{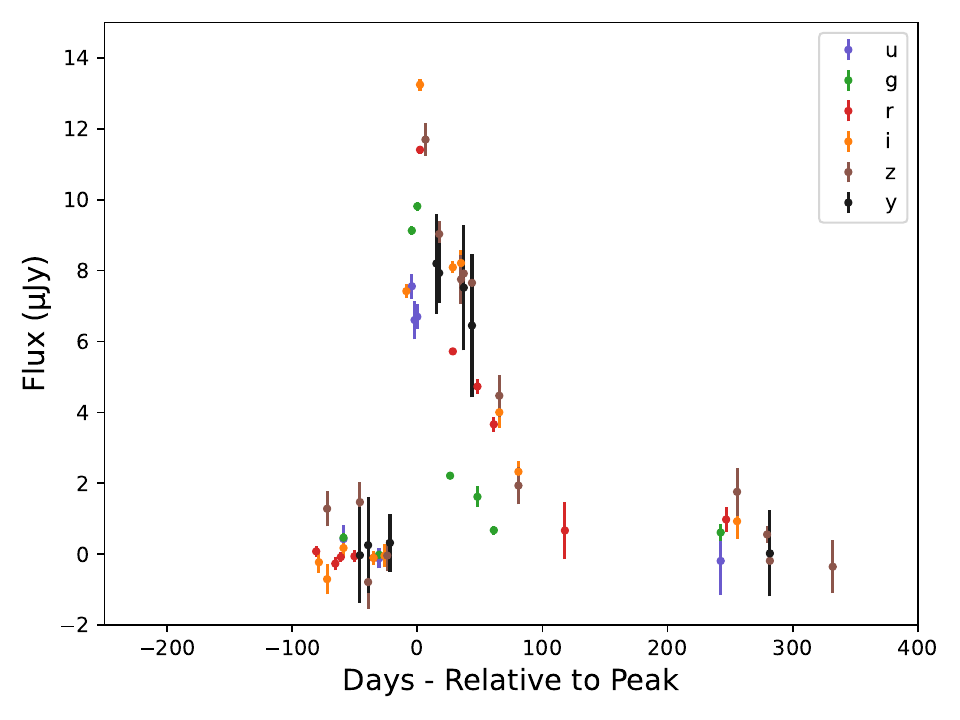}
    \caption{Example of a final simulated LSST lightcurve based on the real ZTF lightcurve for SN II ZTF22abcfics/SN2022ryv. The luminosities have been subjected to some random scatter within measurement uncertainties to mimic the expected scatter from noise in LSST.}
    \label{fig:final_lc}
\end{figure}

Using a normal distribution with width equal to these estimated measurement uncertainties, scatter is added to each lightcurve point. This mimics the expected scatter due to observational noise. This is the final stage in producing a simulated LSST lightcurve from an existing ZTF lightcurve, an example of which is shown in Figure \ref{fig:final_lc}. This process should be replicable for any future photometric survey. All that is required are the lightcurves from one survey, the cadence of the new survey, and the limiting magnitudes for each survey.

\begin{figure}
	\includegraphics[width=\columnwidth]{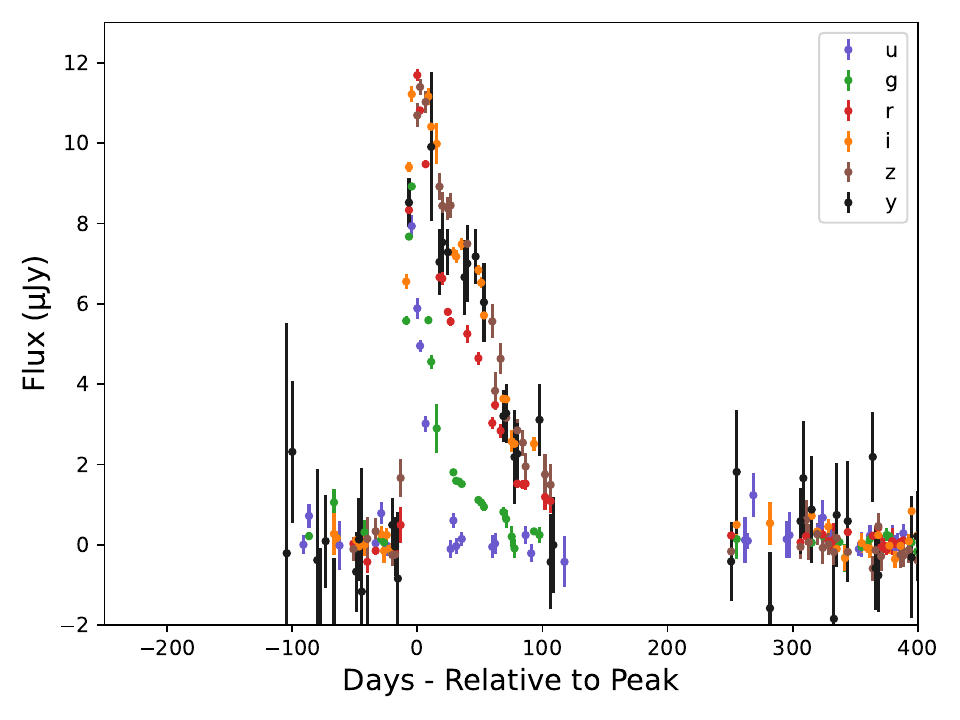}
    \caption{Plot showing a simulated LSST lightcurve based on the real ZTF observations of the SN II ZTF22abcfics/SN2022ryv. This lightcurve has been simulated within the Extended Chandra Deep Field South (ECDFS) and thus given Deep Drilling Field (DDF) cadence, resulting in a much more densely observed lightcurve than in Figure \ref{fig:final_lc}.}
    \label{fig:ddf_cadence}
\end{figure}

Though most of the data from LSST will be obtained in the Wide-Fast-Deep (WFD) fields, through the use of the \texttt{Rubin Survey Simulator}, we can also produce lightcurves with Deep Drilling Field (DDF) cadence. By selecting co-ordinates within one of the deep drilling fields, the simulation will provide visit dates and depths for that position, producing a lightcurve with a much greater cadence. An example is shown in figure \ref{fig:ddf_cadence}. This will occasionally happen naturally in our simulation via the random selection of RA and Dec, or can be artificially forced by selecting a DDF position.

\section{MALLORN Data Set Production}
\label{LSST_Dataset_production}

\subsection{Zwicky Transient Facility Data}

The total number of ZTF objects used to generate simulated LSST lightcurves was 2198. Of this, roughly two-thirds are AGN and just less than a quarter are SNe Ia. The remainder are made up of a variety of other transients. The following types of transients are within the ZTF data set and are used within the production of MALLORN: SN Ia, SN Ia-91T-like, SN Ia-91bg-like, SN Ia-02cx-like, SN Ia-pec, SN Ib, SN Ib/c, SN Ic, SN Ic-BL, SN II, SN IIb, SN IIn, superluminous supernovae (SLSN) Types I and II, TDE and AGN.

Although present in small numbers within the ZTF nuclear transient data set, no galactic transients (such as cataclysmic variables) or variable stars were included within the MALLORN data set. The contextual information LSST and the alert brokers will provide (e.g.~host galaxy associations; \citealt[][]{Sherlock}) in combination with the nature of the lightcurves for such objects should be sufficient to exclude most of these objects - and as such it was unnecessary for us to simulate them within the MALLORN data set.

\begin{figure*}
	\includegraphics[width=15cm]{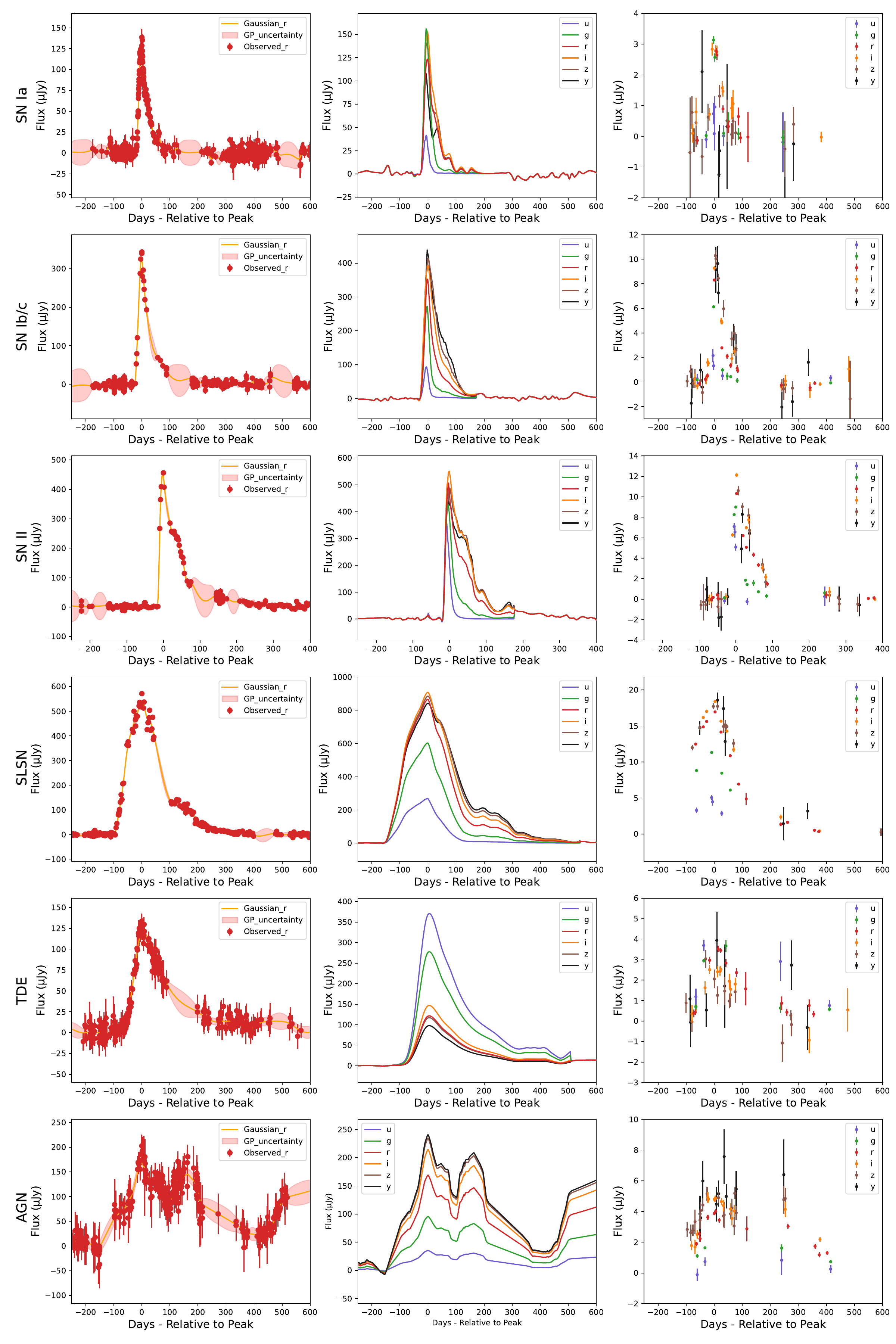}
    \caption{Plots showing the GP fit to the observed data points (LHS), the 6-band continuous SNCosmo plots (centre) and an example of that object simulated at an example LSST luminosity and observing cadence. From top to bottom, the plots are for SN Ia, SN Ibc, SN II, SLSN, TDE and AGN.}
    \label{fig:main_plot}
\end{figure*}

In constructing our simulation, some initial checks on input lightcurve quality were applied. If the lightcurve had less than 30 data points it was screened out. These measures predominantly affected the AGN matched via the \texttt{Milliquas} crosscheck, some of which were close to the ZTF detection limit. This is particularly prevalent for more distant AGN. For other object types, the selection on spectroscopic labels from \texttt{TNS} ensured that most were bright and well detected.

Further safety checks are present in the GP fit to ensure the validity of the fit. The fit will rerun with slightly altered parameters if it fails to reach 90\% of the observed peak flux, repeating until within the range of this threshold value or continuing on after the fifth attempt. This primarily affects narrow lightcurves where the GP might smooth the sharp peak. An MSE value is calculated for each of the GP fits calculated using the method described in \ref{GP_fit}. If the MSE value produced is greater than 0.3, the fit is considered failed and the object is discarded. This was quite a rare occurrence, but was a necessary measure to ensure the validity of the GP fits. From the test of 200 randomly selected objects 85.4\% passed this requirement.

There is an additional safety measure for the DRW model, which excludes AGN with lightcurves having a duration of less than thirty days. The MSE approach was not viable to assess the DRW fit due to the variability in the lightcurves. A large number were visually assessed when testing the DRW fit which verified the validity of the approach. With the stochastic nature of AGN variability and the large number of examples present within the data set, an exact fit to any particular lightcurve was not deemed to be as important as it was for other transient types.

\subsection{SNCosmo Models}
\label{SNCosmo Models}
A variety of models were used to produce the SED models required to generate the colour differences. For thermonuclear supernovae, the `hsiao' model \citep[][]{Hsiao2007} was used for SNe Ia, as it provided good coverage across all six bands. For SNe Ia-91T-like the `nugent-sn91t' model was used \citep[][]{Stern_2004} and for SNe Ia-91bg the `nugent-sn91bg' model was used \citep[][]{Nugent2002}. There is no matching model for SN Ia-02cx-like (or Iax), so the `nugent-sn91t' model was used as an approximate fit for this class. The early-time colours are similar since SNe Iax spectroscopically resemble SNe Ia-91T-like at early times within about a month of explosion \citep{Jha2017}. Similarly for SN Ia-pec (an ill-defined term covering any anomalous SN Ia) there is no matching \texttt{SNCosmo} model and so the `hsiao' model was used to provide an approximate fit for the colours.

For core collapse supernovae, there are a greater variety of models available on \texttt{SNCosmo}. For consistency, it was decided that we would prioritise the `v19' models from \citet{Vincenzi2019}. For each class, there were multiple options to choose from, each based on an individual well-observed event. Where possible, we included multiple models for each type and selected one at random for each MALLORN simulated object, in order to enhance the variety present within the data set. It was necessary to exclude some models from this to ensure the production of quality six-band lightcurves. These models were excluded predominantly due to poor coverage in the $z$ and $y$ bands. For the full \texttt{SNCosmo} model selection details see table~\ref{tab:sn_models}.

%For SNe Ib, the `v19-2004gq', `v19-2007uy' and `v19-2012au' models were chosen. For SN Ib/c, the `v19-2005bf' and `v19-2006ep' models were selected. For SNe Ic, the `v19-2004gt' and `v19-2011bm' models were used. For SNe Ic-BL, the `v19-1998bw', `v19-2002ap' and `v19-2007ru' models were chosen. For SNe II/IIP the models `v19-2009kr', `v19-2013ab', `v19-2013am', `v19-1987a', `v19-2009ib' and `v19-2016x' were selected. For SNe IIb the models `v19-1999dn', `v19-2006t', `v19-2008aq', `v19-2008bo', `v19-2011dh', `v19-2011ei', `v19-2011fu' and `v19-2011hs' were chosen. And for SNe IIn, the models `v19-2006aa' and `v19-2011ht' were used. 

\begin{table}
    \centering
    \caption{\texttt{SNCosmo} models chosen for each transient type.}
    \label{tab:sn_models}
    \begin{tabular}{ll}
        \hline
        \textbf{Transient Type} & \textbf{Selected Models} \\
        \hline
        SN~Ia & hsiao \\
        SN~Ia-91T & nugent-sn91t \\
        SN~Ia-91bg & nugent-sn91bg \\
        SN~Ia-02cx & nugent-sn91t \\
        SN~Ia-pec & hsiao \\
        SN~Ib & v19-2004gq, v19-2007uy, v19-2012au \\
        SN~Ib/c & v19-2005bf, v19-2006ep \\
        SN~Ic & v19-2004gt, v19-2011bm \\
        SN~Ic-BL & v19-1998bw, v19-2002ap, v19-2007ru \\
        SN~II/IIP & v19-1987a, v19-2009ib, v19-2009kr, \\
                  & v19-2013ab, v19-2013am, v19-2016x \\
        SN~IIb & v19-1999dn, v19-2006t, v19-2008aq, v19-2008bo, \\
                & v19-2011dh, v19-2011ei, v19-2011fu, v19-2011hs \\
        SN~IIn & v19-2006aa, v19-2011ht \\
        SLSN-II & v19-2011ht \\
        SLSN-I & Linear Temperature Blackbody \\
                & \citep[][]{Nicholl2017,Gomez2024}\\
        TDE & Linear Temperature Blackbody \citep[][]{VanVelzen2021}\\
        AGN & Constant Temperature Blackbody \\
        \hline
    \end{tabular}
\end{table}

There is no model or viable appropriate substitute within \texttt{SNCosmo} for TDEs or AGN. For these it was necessary to construct blackbody models -- an excellent approximation for TDEs in particular. For TDEs a time-evolving blackbody model with a linear temperature change was implemented. A TDE peak temperature is selected at random from a list of ZTF-based temperatures. The average of the TDE temperature distribution is $1.3 \times 10^4$\,K with a standard deviation of $5 \times 10^3$\,K and a maximum temperature of $3.2 \times 10^4$\,K. Following \citet[][]{VanVelzen2021}, a linear temperature evolution list was produced via multi-band modelling of the $g$ and $r$ bands of the ZTF TDEs, supplemented by UV data from other surveys if available\footnote{\url{https://github.com/sjoertvv/manyTDE}}.
The value for the linear change in temperature over time (K/day) which most closely matches the selected peak temperature is then chosen from the list of ZTF-based temperatures. This linear temperature change, which can be positive or negative, is then applied to the temperature used in the model for 100 days after the peak date of the lightcurve. After this 100 days a constant temperature is used. A maximum temperature of 40000\,K and minimum temperature of 8000\,K were implemented to prevent the linear temperature change from producing unphysical TDEs. While we could have used the best-fit temperature and evolution for the ZTF TDEs, randomly selecting a temperature for each simulated TDE was considered preferable as it treats all of the objects the same and helps to enhance the variety within the data set.

For AGN, a constant-temperature blackbody model is used. AGN spectra are traditionally approximated with a power-law \citep[][]{Urry1995}, however a blackbody also provides a reasonable approximation in the optical range.  The AGN temperatures are produced via the same multi-band modelling method as the TDE temperatures using ZTF AGN, and are again selected at random to enhance variety. The average of the AGN temperature distribution is $1.5 \times 10^4$\,K with a standard deviation of $2.3 \times 10^4$\,K, with the standard deviation inflated by some blue outlier AGN.

The other classes with no built-in examples present in \texttt{SNCosmo} are the SLSNe. For SLSNe-I, we used a blackbody model with UV absorption below 3000\,\AA, based on the magnetar model described in \citet[][]{Nicholl2017}, coupled with a linear temperature change similar to our TDEs. For this model, a random start and end temperature are selected (from between 10000K-15000K and 5000K-8000K respectively) based on the values in \citet[][]{Gomez2024}. A cooling rate (K/day) is then calculated by dividing the difference between these temperatures by a randomly selected cooling time (30-100 days). This linear temperature change is then applied for 70 days after peak. For SLSNe-II the brightest SN IIn model, `v19-2011ht', was selected, as SLSNe-II are thought to be the high end luminosity tail of SNe IIn \citep[][]{GalYam2019}.

\subsection{Large Scale Production}
\label{Large Scale Production}
For each real ZTF object within the source data set, it was possible to produce several distinct LSST lightcurves, enabling us to upsample the data set. 

\begin{figure*}
	\includegraphics[width=15cm]{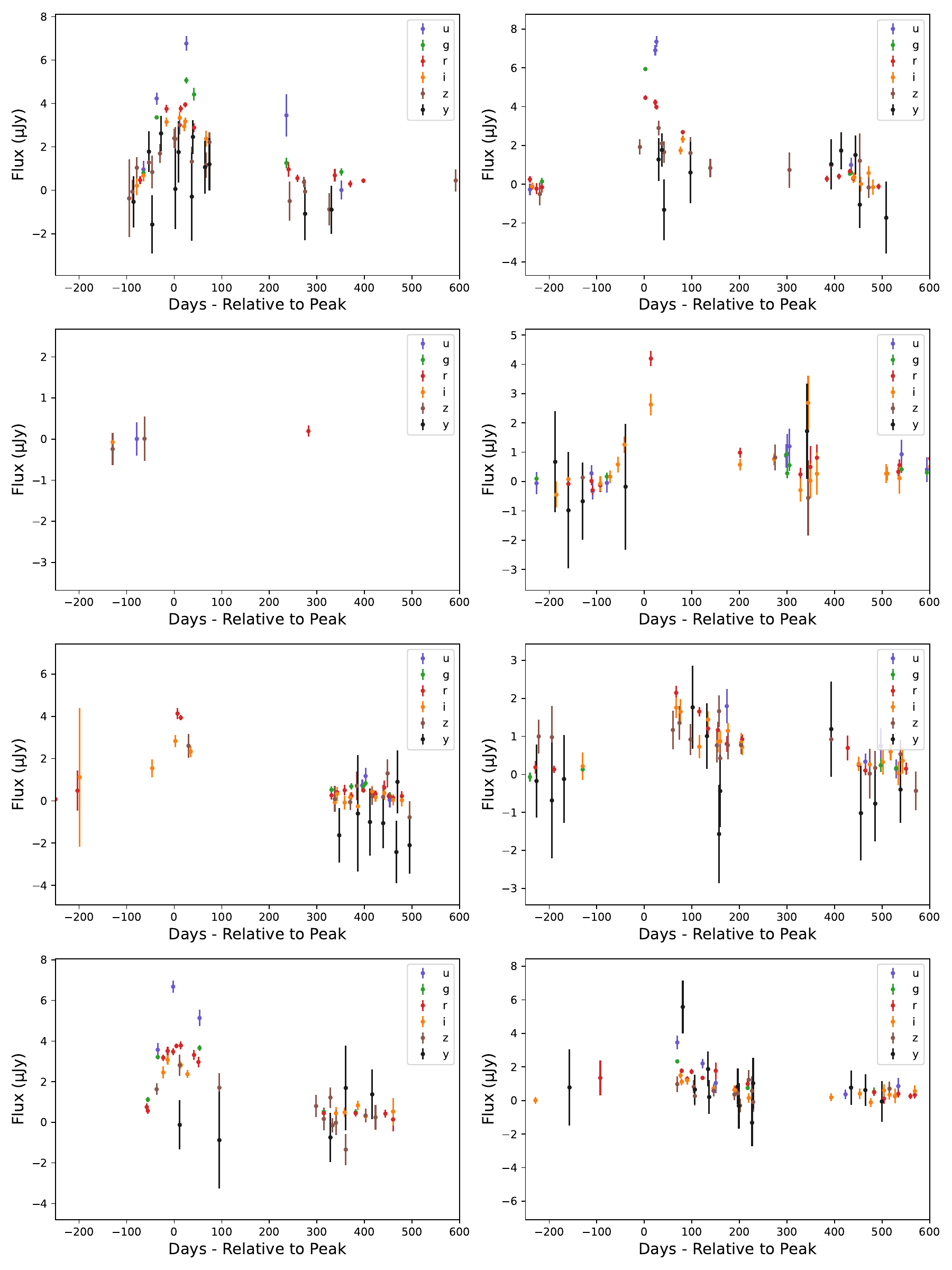}
    \caption{Eight simulated LSST lightcurves produced from the same real ZTF object (ZTF21abcgnqn). The LSST cadence and the scatter parameters used to generate the lightcurves provide sufficient variety to allow for multiple distinct lightcurves to be created from the same ZTF object.}
    \label{fig:cadence_proof}
\end{figure*}

The $\sigma$ and $\delta$ parameters described in section \ref{rescaling} provide variation in the peak absolute and apparent magnitudes. Using a standard deviation of 0.3 for $\delta$ provided a good compromise between adding variation and avoiding excessively bright or faint transients. We get variety in colour via using different templates for core-collapse SNe and different temperatures for TDEs, AGN and SLSNe. The LSST cadence supplies sufficient variety that objects produced from the same source with a different position on the sky and at a different date are distinct from each other as shown in Figure \ref{fig:cadence_proof}.

Through the combination of the random scatter component in the distance calculation and the peak magnitude scatter component, it was possible for objects to be generated at luminosities that were not viable for detection. To prevent these objects from leaking into the MALLORN data set, a safety measure was implemented which removed objects that peaked below the LSST limiting magnitude. This mostly excluded AGN whose source objects were already close to the ZTF limit.

This allowed for the MALLORN data set to reach a potential total of roughly 20000 objects. Due to the quality cuts a final total of 10178 lightcurves were produced and included in the MALLORN data set.

Determining the redshifts of the observed objects is vital for many of the science goals for LSST. Ideally the redshift is determined spectroscopically, however this is not possible to obtain for all galaxies in LSST or even just all transient hosts. LSST instead aims to use its broad-band photometry and predictions from galaxy spectral energy distributions to determine a photometric redshift value. This approach allows for a redshift value to be determined for a much larger quantity of objects, however it is less accurate than a spectroscopic redshift \citep[][]{Graham2017, Jones2024}. For the objects in the MALLORN data set a simulated photometric redshift is created. The `true' simulated redshift is known for all objects. The uncertainty in LSST photometric redshifts is expected to follow \citet{Graham2017}:

\begin{equation}
    z_{\text{err}} = 0.02 \times (1 + z)
\end{equation}
A simulated photometric redshift for each MALLORN object is drawn from a normal distribution centred on the real redshift value with a standard deviation equal to $z_{\rm err}$. This approximates the photometric redshift and its uncertainty that will be available in LSST data releases, and may be an important feature in transient photometric classification.

Storing our simulated objects, and associated classifications with lightcurves, requires a unique identifier for each one. To produce a unique object ID for each of the simulated lightcurves produced, three Sindarin words are chosen at random and combined. Sindarin is a fictional language J.R.R. Tolkien developed for his Elves in \textit{`The Lord of the Rings'} \citep[][]{Tolkien1954}, which felt appropriate given the data set is named after the golden-leaved mallorn trees of Lothl\'orien. The Sindarin dictionary\footnote{Sindarin Dictionary - \url{https://www.ambar-eldaron.com/english/downloads/sindarin-english.pdf}} chosen contained 375 words, therefore allowing for roughly $5\times10^7$ possible combinations which was more than sufficient for the roughly $2\times10^4$ lightcurves generated for the MALLORN data set. The simulations were run on the Queen's University Belfast high-performance computing facility Kelvin2.

\section{Data Challenge Structure}
\label{Data Challenge Structure}
The MALLORN Astronomical Classification Challenge is hosted on \texttt{Kaggle}\footnote{\url{https://www.kaggle.com/}}. This site has been used previously for other astronomical classifier challenges \citep[][]{Kitching2011,Harvey2014,Kessler2019}, providing the opportunity for both astronomers and members of the general public to participate in producing and testing the effectiveness of classifiers. 

For the purposes of the classifier challenge the data set is split into training set, public testing set and private testing set files. For the training set, the spectral type and true redshift are known, thereby providing all the information that would be required to train a machine learning classifier (or other alternative) on that portion of the MALLORN data. This comprises 30\% of the overall data. The remaining 70\% is the testing set, on which the responses to the classifier challenge will be judged. For this portion of the data, the user will not know the spectral type and only an estimated photometric redshift is provided. The testing set is split into two portions - public and private. For the public testing set the user will be able to see the overall performance of their classifier when they submit, therefore allowing them to make changes to improve their classifier and resubmit another response to the classifier challenge. For the private testing set, the user's performance is not shown to them. The purpose of this is to prevent a user from over-fitting by iteratively training to improve their score only on the public test set. The inclusion of a private testing set keeps the classifier challenge fair and helps to ensure that the classifiers tested on it will be effective at classifying actual LSST TDEs and not just fine-tuned to spot the TDEs included in the MALLORN public testing data set.

The lightcurve files contain information on the object ID, date (MJD), flux ($\mu$Jy), flux error ($\mu$Jy) and filter. Multiple objects are assigned to one lightcurve file. There are training set and testing set log files which contain information on the object ID, redshift, redshift error, extinction coefficient $E(B-V)$, spectral type, and an English translation providing the meaning of the object ID. The training set log is set to have spectroscopically determined redshifts, and as such has the actual simulated redshift value and no error value. Additionally, the spectral type is available for the training set. For the testing sets, the photometrically determined redshifts and their corresponding error values are used and the spectral type is not included. The $E(B-V)$ value is included to allow for the lightcurves to be de-extincted easily by the user. It was chosen to include these values directly rather than the RA and Dec to simplify the process of handling the data for non-astronomers who may wish to participate in the MALLORN classifier challenge. There are additional master log files detailing other parameters generated in production which are kept private.

To further lower the minimum barrier to entry for the classifier challenge, the Using MALLORN Data Google Colab Notebook\footnote{\url{https://colab.research.google.com/drive/1N7Q1bxc2gxBuOv2eD3fTYsrxoLC2dQAP}} details the process of loading in and de-extincting a lightcurve from the files and loading the relevant information for that lightcurve from the log files. The purpose of this notebook is primarily to lower the barrier of entry for non-astronomers and better enable their participation in the classifier challenge. This is of particular importance as there may be many who are talented in machine learning and computer science who could otherwise be put off from the challenge by its inaccessibility. Of the top three best performing entries to PLAsTiCC, only the top entry, detailed in \citet[][]{Boone2019}, was from an astronomer, with the other top performing entries from members of the general public \citep[][]{Hlozek2023}. We aim to make this classifier challenge as accessible as possible to maximise the potential gains in classifier performance prior to the start of LSST and therefore bolster our ability to identify TDEs with LSST.

Users are welcome to identify other types of transient within the data (and some participants may be specifically interested in learning how to find other classes in LSST), but the final score in the classifier challenge will only be based on the ability of the classifier to identify TDEs within the MALLORN data set. Responses will be evaluated using the F1 score \citep[][]{Powers2011}. This is a common machine learning evaluation metric which assesses the precision and recall scores of a model. Precision is the accuracy of the model, evaluated based on the ratio of true positive values retrieved and the total number of positive values retrieved. Recall is the sensitivity of the model, determined from the ratio between the number of correctly predicted positive values and the total number of positive values in the data set. The F1 score itself can be interpreted as the harmonic mean of these two values, with an F1 score of 1 corresponding to the best possible performance and a score of 0 to the worst. The precision and recall values have equal weighting in the calculation of the F1 score. The F1 score is calculated using the following formula:
\begin{equation}
    \mathrm{F} 1=\frac{2 \times \mathrm{TP}}{2 \times \mathrm{TP}+\mathrm{FP}+\mathrm{FN}}
\end{equation}

In the above equation TP is the number of true positives (correct positive predictions), FN is the number of false negatives (incorrect negative predictions) and FP is the number of false positives (incorrect positive predictions). The competitor submitted classifications will be compared against the known types from the data production log files to determine the F1 score. Using the F1 score to select the most successful model disfavours models that over-predict the amount of TDEs in order to inflate their completeness. Whichever competitor achieves the highest F1 score will be determined to be the winner of the MALLORN Classification Challenge. The MALLORN Astronomical Classification Challenge is live as of the time of writing, having launched on 15/10/2025, and closes on 30/01/2026.

The competitor who produces the highest scoring classifier will win a \texteuro1000 prize. Whilst second and third place will receive \texteuro250 and \texteuro100 respectively. In addition to the monetary reward the best performing competitors will be invited to add their classifier to a platform which is being developed to host the best TDE classifiers. The benefit of this being that their classifier, along with others, will contribute to and receive credit for many future TDE discoveries.

\section{Discussion \& Conclusions}

Having produced our simulated data set of ~10,000 LSST transients, we now assess what their lightcurves imply for transient science, and particularly TDE science. During the creation of the MALLORN data set it became clear that some LSST bands will be much more useful for transient discovery than others, largely due to the differences in detection limits and the cadence strategy chosen. The $g$, $r$ and $i$ bands will be the most useful, with frequent observations allowing for effective analysis of time evolution and detection limits which are capable of providing good signal-to-noise detections of fainter transients. 

The $u$ band has a sufficient detection limit to provide solid detections for fainter transients. This band is expected to be particularly important for the photometric identification of TDEs, which are inherently blue \citep[][]{Gezari2021}. Supernovae have been shown to peak at much redder wavelengths with increasing redshift, whereas TDEs continue to peak at blue wavelengths with increasing redshift \citep[][]{VanVelzen2021}. Therefore, these $u$ band observations will likely play a key role in the photometric identification of TDEs. The utility of this band is however limited by its less-frequent use in the LSST cadence.

For the $z$ \& $y$ bands, the combination of infrequent sampling and shallower detection limits significantly restrict their utility for transient discovery and likely for photometric classification. This is particularly prominent for $y$ band observations of fainter transients, which are much noisier than in other bands (see e.g. Figures \ref{fig:final_lc} \& \ref{fig:main_plot}). For much brighter objects, these filters will likely provide good IR coverage of objects, but they will be of very limited use in the discovery and photometric classification of fainter transients.

We analyse the likelihood of detecting TDEs in LSST using our MALLORN data set and the metrics outlined in \citet[][]{BucarBricman2023}. 
%The \texttt{prepeak} metric requires at least two detections prior to the peak of the lightcurve. 
The \texttt{some\_color} metric requires at least one detection pre-peak ($t<t_{\text{peak}} - 10$ days), at least three detections in at least three different filters within 10 days of peak, and at least two detections in at least two different filters 10-30 days after peak. The \texttt{some\_color\_pu} (`plus $u$') metric requires at least one detection pre-peak ($t<t_{\text{peak}} - 10$ days), at least one detection in the $u$ band and at least one detection in another filter within 10 days of the peak and at least one detection in another filter 10-30 days after peak. These criteria give a rough estimate of whether a TDE can be selected photometrically, and whether there is enough information to extract physical insight. A detection within 10 days of the peak is required for a reliable estimate of the black hole mass \citep[][]{BucarBricman2023}. Detections within 10-30 days after peak are necessary to estimate the fallback rate and hence the mass and impact parameter of the star which produced the TDE \citep[][]{GuillochonRamirezRuiz2013}.

Using a simulated population of TDEs (1000 at varying redshifts, based on a "normal" TDE, PS1-10jh \citep[][]{Gezari2012}, and a "fast and faint" TDE, iPTF16fnl \citep[][]{Blagorodnova2017}) and \texttt{Baseline v1.5}, \citet[][]{BucarBricman2023} evaluated how many TDEs would pass these detection metrics in LSST. 7.5\% of TDEs were found to pass the \texttt{some\_color} metric and 1.5\% were found to pass the \texttt{some\_color\_pu} metric. These results were found to be heavily dependent on redshift, with mainly low-redshift TDEs passing the metrics, particularly in the case of the "fast and faint" TDEs. 

\begin{figure}
	\includegraphics[width=\columnwidth]{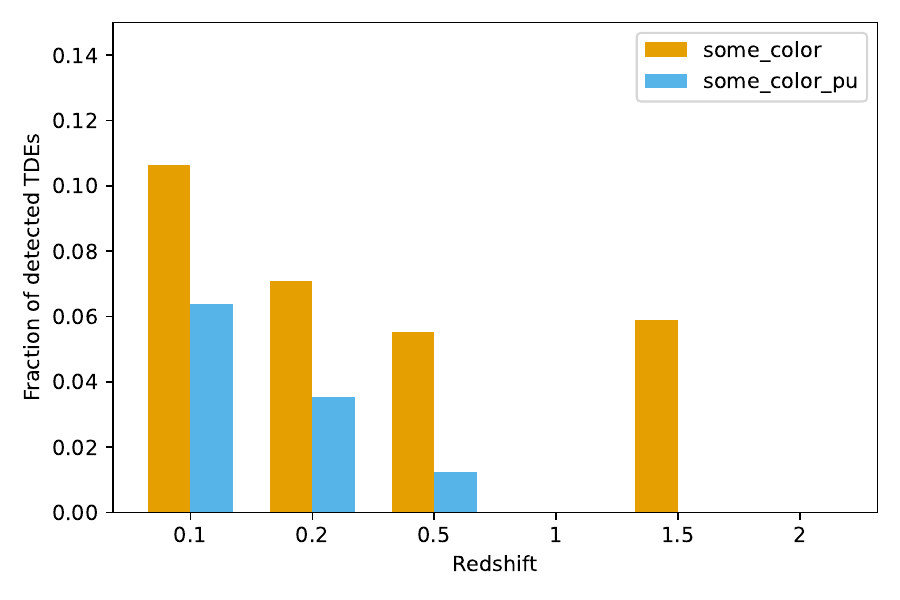}
    \caption{Bar chart showing the fraction of TDEs in MALLORN to pass the \texttt{some\_color} and \texttt{some\_color\_pu} metrics from \citet[][]{BucarBricman2023}. The fraction of TDEs able to pass both metrics is shown to decrease with increasing redshift, particularly in the case of the \texttt{come\_color\_pu} metric.}
    \label{fig:tde_analysis}
\end{figure}

We recalculate these metrics for MALLORN to update the predictions for TDE performance, as our simulations included a much more representative set of TDEs and the latest LSST cadence. Defining a successful detection as a $5\sigma$ observation, 6.5\% of the TDEs in MALLORN pass the \texttt{some\_color} metric whilst 2.7\% pass the \texttt{some\_color\_pu} metric. The results are broadly similar, with the updated version of the baseline accounting for some of the discrepancy. Additionally, the measures taken to ensure a minimum level of quality in MALLORN may slightly inflate the amount of TDEs which pass the metric in comparison to \citet[][]{BucarBricman2023}. However, it is worth noting that our simulation largely ignores low-redshift TDEs due to how we rescale the ZTF objects. Including more TDEs at $z<<0.1$ (though a small fraction of LSST TDEs) would somewhat increase these metrics.
A correlation between redshift and metric performance was also observed in MALLORN, as shown in figure \ref{fig:tde_analysis}. The fraction of TDEs which pass both metrics decreases with increasing redshift, particularly in the case of the \texttt{some\_color\_pu} metric, which has no successes at $z\gtrsim1$.

There are various types of transients (such as kilonovae and SNe Ibn) not included within the MALLORN data set, as they were not present in the ZTF nuclear transient data set that was used to generate it. Given that these types of transients did not appear in the extensive ZTF sample, they are clearly rare in galaxy nuclei and so we do not expect them to be a significant contaminant in the search for TDEs. Ambiguous Nuclear Transients \citep[ANTs;][]{Wiseman2025} are also not included as a class in the MALLORN data set, due to their rarity. However, our data set does include some TDEs which occurred in AGN.

There are some limitations of the MALLORN data set that could be improved upon if this method was to be repeated for a future survey. Many of the \texttt{SNCosmo} models lack good SED model coverage of the $i$, $z$ \& $y$ bands. By carefully selecting models we were able to create reasonable colour curves and k-corrections, but improved coverage in redder bands (which LSST will provide) would allow for this to be more accurate in future. Bright and nearby objects that LSST observes will likely have additional behaviour prior to the rise visible in the ZTF data and potentially more activity after the lightcurve fades to the ZTF detection limit. The MALLORN data set focused on being grounded in real observations, and consequently we avoided extrapolating to produce the simulated data required to produce these lightcurves. We believe that these brighter transients should be easier to classify, whereas the fainter objects we are focusing on will be more difficult. Therefore our data set will be more challenging for current classifiers and provide a greater avenue for improvement in advance of the start of LSST.

The selection of temperature at random rather than matching the observed temperatures for all chosen objects may result in an inaccurate distribution of temperature in the TDEs generated in MALLORN. This would be the case if there is a strong correlation between temperature and luminosity, in which case we may inaccurately simulate some cooler distant TDEs that would be harder to classify. 
There is some evidence for a correlation between luminosity and temperature, but with a very large intrinsic scatter that suggests this effect will be modest within our data set \citep[][]{Yao2023}.
Despite this potential issue, this approach was still necessary in order to produce multiple simulated objects from one source object. The linear temperature change over time is not randomly selected, the corresponding value for the nearest peak temperature is selected to ensure physically reasonable temperature distributions.

The class imbalances in time-domain transient astronomy present some inherent challenges, which may make classification difficult -- especially for some traditional computer science approaches. A simulated data set grounded in observational reality helps to overcome this issue, and is therefore beneficial to both fields, providing a strong example of the benefit such interdisciplinary approaches can create.

%[Brief discussion of computational intensity of the approaches employed by competitors. If two approaches produce broadly comparable results, but one uses significantly less resources, that approach should be deemed preferable by the community. Unsure that there's actually a way to implement that into the data challenge but still worth mentioning.] - leave for next paper

The MALLORN data set and Astronomical Classification Challenge will allow us to better prepare in advance of LSST, and contribute to a more effective and immediate start to transient research once the survey begins. The method developed for the production of MALLORN is highly replicable. It is possible to use it to generate simulated data for any future photometric survey from a previous one, given the limiting magnitudes of both and an expected cadence. By publicly providing our method via Google Colab (and releasing on GitHub after the challenge) we expect that this could be useful to prepare for other future surveys, such as the \textit{Roman} High-Latitude Time-Domain Survey.

\section*{Acknowledgements}

Thanks to M. Kowalski and S. Reusch for running the \texttt{AMPEL} Nuclear Filter and creating the ZTF nuclear transient data set used to create the MALLORN data set.
DM acknowledges a studentship funded by the Leverhulme Interdisciplinary Network on Algorithmic Solutions.
MN is supported by the European Research Council (ERC) under the European Union’s Horizon 2020 research and innovation programme (grant agreement No.~948381).

%%%%%%%%%%%%%%%%%%%%%%%%%%%%%%%%%%%%%%%%%%%%%%%%%%
\section*{Data Availability}

%The inclusion of a Data Availability Statement is a requirement for articles published in RASTI. Data Availability Statements provide a standardised format for readers to understand the availability of data underlying the research results described in the article. The statement may refer to original data generated in the course of the study or to third-party data analysed in the article. The statement should describe and provide means of access, where possible, by linking to the data or providing the required accession numbers for the relevant databases or DOIs.

All the data produced as part of this work has been made publicly available via Kaggle (\url{https://www.kaggle.com/competitions/mallorn-astronomical-classification-challenge/overview}). The spectroscopic types of the private data set will remain hidden during the duration of the data challenge and will be made available upon its completion. Google Colab notebooks have been made available that describe how the data was produced (\url{https://colab.research.google.com/drive/1oy96r29Zs4U5Hl-THsZPCnOQbuz21hl5}) and how to use the data (\url{https://colab.research.google.com/drive/1N7Q1bxc2gxBuOv2eD3fTYsrxoLC2dQAP}).



\bibliographystyle{mnras}
\bibliography{mallorn} % if your bibtex file is called example.bib

%\begin{thebibliography}{99}
%\end{thebibliography}

%%%%%%%%%%%%%%%%%%%%%%%%%%%%%%%%%%%%%%%%%%%%%%%%%%

%%%%%%%%%%%%%%%%% APPENDICES %%%%%%%%%%%%%%%%%%%%%

%\appendix



%%%%%%%%%%%%%%%%%%%%%%%%%%%%%%%%%%%%%%%%%%%%%%%%%%

% Don't change these lines
\bsp	% typesetting comment
\label{lastpage}
\end{document}